\documentclass[11pt,a4paper,english,nofootinbib,,superscriptaddress]{revtex4}
\usepackage{lmodern}

\usepackage[T1]{fontenc}
\usepackage[latin9]{inputenc}
\usepackage{babel}
\usepackage{amsmath}
\usepackage{graphicx}
\usepackage{amssymb}
\usepackage{esint}
\usepackage[unicode=true, pdfusetitle,
 bookmarks=true,bookmarksnumbered=false,bookmarksopen=false,
 breaklinks=false,pdfborder={0 0 1},backref=false,colorlinks=false]
 {hyperref}
\usepackage{amsmath}
\usepackage{amssymb}
\def\b{\begin{equation}}
 \def\e{\end{equation}}

\makeatletter
\usepackage{latexsym}\usepackage{bm}

\makeatother

\begin{document}

\title{Particle Creation and Excited-de Sitter Modes}

\author{M. Mohsenzadeh}
\email{mohsenzadeh@qom-iau.ac.ir}
\affiliation{Department of Physics, Qom Branch, Islamic Azad University, Qom, Iran}
\author{E. Yusofi}
\email{e.yusofi@iauamol.ac.ir}
\affiliation{Department of Physics, Ayatollah Amoli
Science and Research Branch, Islamic Azad University, Amol, Mazandaran, Iran}

\date{\today}

\begin{abstract}

\noindent \hspace{0.35cm}
Recently, in Ref.\cite{moh1}, we introduced exited-de Sitter modes to study the power spectrum which was finite in Krein space quantization and the trans-Plankian corrections due to the exited modes were non-linear. It was shown that the de Sitter limit of corrections reduces to what obtained via the several previous conventional methods, moreover, with such modes the space-time symmetry becomes manifest. In this paper, inspired by Krein method and using exited-de Sitter modes as the fundamental initial states during the inflation, we calculate particle creation in the spatially flat Robertson-Walker space-time. It is shown that in de Sitter and Minkowski space-time in the far past time limit, our results coincides to the standard results.\\

\textbf{Pacs}:04.62.+v, 04.25.-g, 98.80.Cq\\

\end{abstract}

\maketitle

\section{Introduction and Motivations}
A remarkable and unique property of the inflationary scenario of  early Universe is that it opens a rapturous possibility to directly observe the consequence of a genuine quantum-gravitational effect, an important one is the production of quasi-classical fluctuations of quantum fields in strong external gravitational fields. Generally, this effect was known as particle production \cite{bri}, however, it is obvious at present that what can be measured are in fact not particles but rather inhomogeneous fluctuations (perturbations) of the gravitational fields. It seems that in the non-inflationary cosmological models, the effect of particle production is exceedingly small and does not lead to observable subsequence. But just the opposite, in the simplest versions of the inflationary theory \cite{haw}, not only the minimum perturbations of the gravitational potential generated can be sufficient to explain the large-scale structure in the Universe and the galaxy formation, but also the Gaussianity of statistics and their predicted spectrum (that is approximately flat) have been successfully quantitatively verified by the COBE discovery \cite{smo}.\\
The problem of particle production due to a curved background space-time has a rich history dated back to the early work of Schr\"odinger in 1939 \cite{sch}, later this developed by Parker in the late 60s \cite{par}. Almost all initial studies of the particle production in cosmology in general and also in the inflationary scenario in particular were accomplished using the Heisenberg method (see for instance \cite{zel} and \cite{sta}). A characteristic feature of this method is using of the mode functions that satisfying a classical wave equation and also the Bogoliubov transformations for the annihilation and creation operators. \\
On the other hand, in spite of many success stories in quantum field theory in curved space-time and also in the inflationary scenario, the concept of particle and vacuum in generic curved space-time still suffers from some significant pitfalls. One knows that the inflation could be described by almost de Sitter space-time \cite{linde}. In flat space-time, there is a well-defined vacuum state, but in curved space-time, the concept of vacuum is not very obvious and exists obscurity in the choice of vacuum. If we consider universe as the exact de Sitter space-time during the inflation, there exists a actual class of vacuum states invariant under the symmetry group of the de Sitter space-time. However, the inflating universe is not exact de Sitter space-time, however, in the first approximation, it may be described by de Sitter space-time. Also, recent Planck results as the observational motivation, motivated us to use non-Bunch-Davies vacuum. Therefore, in this work, we consider \emph{excited- de Sitter} mode which was introduced in \cite{moh1,yus} instead of Bunch-Davies (BD) mode as the primary modes during the inflation to study the issue of the particle creation. In \cite{yus}, such non-trivial modes have been used to calculate the non-linear trans-Planckian corrections of power spectrum, and, by taking into account the effects of trans-Planckian physics and these alternative modes, we calculated the spectra in the standard approach which leads to a finite power spectrum and the corrections were non-linear noting that de Sitter limit of corrections reduce to the linear form that obtained from the several previous conventional methods.

The study of particle creation can be treated in the  early universe by a number of mechanisms \cite{kol}. In this work, inspired by the method of removing effects of the background space-time in a curved space-time in general and also the definition of two-point function in Krein space in particular, we will offer a definition for spectrum of created particles which caused by the perturbations of purely geometrical and can be interpreted as the perturbation of the geometry. Also, one of the major problems involved in the process of particle creation in a curved space-time is actually how to define the initial vacuum state \cite{bri}. In the present paper by using the result of Ref.s\cite{moh1,yus}, we consider the particle creation from the vacuum fluctuation during the inflation, when the mode functions represent an asymptotically mode function. In this approach, the de Sitter background has been chosen and accordingly the particle production can be studied explicitly through the solution of the wave equation for a scalar field in the approximate de Sitter space-time.

The organization of the paper is as follows: In Sec. 2 we review the quantization in expanding background of scalar field during the inflation. In the main Sec. 3, we first review the power spectrum in Krein space with excited-de Sitter mode and then calculate the particle production in de Sitter background space-time, by using excited-de Sitter mode as the fundamental initial states during the inflation. Conclusions and outlook are given in the final section.

\section{QUANTIZATION IN EXPANDING BACKGROUNDS}
Quantization of a scalar field in curved space-time follows in straight analogy with the quantization in flat Minkowski space-time where the gravitational metric behaves as a classical external field which could be in general non-homogeneous and non-stationary. The following metric is commonly used to describe the universe during the inflation\footnote{Note we use the units in which $ c =G = \hbar =1$.}: \b \label{equ1}
ds^{2}=dt^{2}-a(t)^{2}{d\textbf{x}}^{2}=a(\eta)^{2}({d\eta}^2-{d\textbf{x}}^2),
\e
where for de Sitter space, the scale factor is given by $a(t)=\exp(Ht)$ and
$a(\eta)=-\frac{1}{H\eta}$. Noting that $\eta$ is the conformal time and $H$ is
the Hubble constant. There are some models of inflation but the
popular and simple one is the single field inflation in which a
minimally coupled scalar field (inflaton) is considered in inflating
background
\b \label{equ2}
S=\frac{1}{2}\int d^4x\sqrt{-g}\Big(R-(\nabla \phi)^2-m^2\phi^2\Big).
\e
The corresponding inflaton field
equation in Fourier space is given by:
\b \label{equ3}
{\phi''}_{k}-\frac{2}{\eta}{\phi'}_{k}+(k^{2}+a^2m^2)\phi_{k}=0,
 \e
where prim is the derivative with respect to conformal time $\eta$.
For the massless case, with the rescaling of $u_{k}=a\phi_{k},$ equation (\ref{equ3}) becomes
\b
 \label{equ4} {u''}_{k}+\omega_{k}^{2}(\eta)u_{k}=0,
  \e
with $ \omega_{k}^{2}(\eta)=k^{2}-\frac{a''}{a} $. The solutions of the above equation give the negative and positive frequency modes. The general solutions of this equation can be written as
\cite{arm}:
\b \label{equ5}
u_{k}=A_{k}H_{\mu}^{(1)}(|k\eta|)+B_{k}H_{\mu}^{(2)}(|k\eta|),
 \e
where $ H_{\mu} $ are the Hankel functions of the first and second kind \cite{nal}. The canonical quantization can be carried out for the scalar field $ u $ and its canonically conjugate momentum $ \pi\equiv u' $, by imposing the equal-time commutation relations, namely $ [u(x,\eta),\pi(y,\eta)]=i\delta(x-y) $, and also by implementing secondary quantization in the Fock representation. After convenient Bogoliubov transformations, one obtains the transition amplitudes for the vacuum state and the associated spectrum of the produced particles in a non-stationary background \cite{muk}.

Motivated by this fact that the inflation starts in approximate-de Sitter space-time, we offer an excited-de Sitter solution as the fundamental mode during the inflation that asymptotically approaches to de Sitter mode as follows \cite{yus}

$$
u_{k}=\frac{A_{k}}{\sqrt{k}}\Big(1-\frac{i}{k\eta}-\frac{1}{2}(\frac{i}{k\eta})^{2}\Big)e^{-ik\eta} $$
\b \label{equ6}
+\frac{B_{k}}{\sqrt{k}}
\Big(1+\frac{i}{k\eta}-\frac{1}{2}(\frac{-i}{k\eta})^{2}\Big)e^{ik\eta},
\e
noting that up to second order this excited-de Sitter mode reduces to the de Sitter mode

\b \label{equ7}
u^{dS}_{k}=\frac{A_{k}}{\sqrt{k}}\Big(1-\frac{i}{k\eta}\Big)e^{-ik\eta}+\frac{B_{k}}{\sqrt{k}}
\Big(1+\frac{i}{k\eta}\Big)e^{ik\eta}.
\e
For the mode equation (\ref{equ4}), the positive frequency solution is given by \cite{moh1}
\b \label{equ8}
u_{k}^{exc}\simeq\frac{1}{\sqrt{k}}\Big(1-\frac{i}{k\eta}-\frac{1}{2}(\frac{i}{k\eta})^{2}\Big)e^{-ik\eta}.
\e
 If we consider $(\frac{i}{k\eta})^{2}\rightarrow0$ for the far past time, the above mode function (\ref{equ8}) leads to the exact BD mode:
  \begin{equation}
 \label{equ9} u_{k}^{BD}=\frac{1}{\sqrt{k}}(1-\frac{i}{k\tau})e^{-ik\tau}.
\end{equation}
In this case, we have $a(t)=e^{Ht}$, or $ a(\tau)=-\frac{1}{{H}\tau}$, with $H =constant$ for  early universe.
\section{Calculations with Excited-de Sitter Mode}
\subsection{Power Spectrum}
As one knows the vacuum expectation value of the energy-momentum tensor becomes infinite which by the means of the normal ordering one gets a finite value in the free field theory in flat space-time, however, in the case of curved space-time, following remedy is usually used (equ. (4.5) in \cite{bri}),
\begin{equation}
\label{equ11} \langle\Omega|:T_{\mu\nu}:|\Omega\rangle=\langle\Omega|T_{\mu\nu}|\Omega\rangle-\langle0|T_{\mu\nu}|0\rangle,
 \end{equation}
where $ |\Omega\rangle $ and $ |0\rangle $ stand for the vacuum states of the theory and flat space-time, respectively. One may interpret the mines sign as the effect of the background. This technique means that one can remove the effects of the background space-time. This interpretation of removing the related effects of a space-time, resembles Krein space approach \cite{Gazea,moh2}, where the renormalization method is accomplished by the support of the negative norm solutions of the wave equation and therefore the minus sign in (\ref{equ11}) appears because of the negative norm solutions. Also, in Krein space, the two-point function is defined by\cite{moh1}:
\begin{equation}
\label{equ12}
\langle\Omega^{Krein}|{:\phi^{2}:}|\Omega^{Krein}\rangle=\langle
\phi^{2}\rangle_P+\langle \phi^{2}\rangle_N, \end{equation}
where the subscript $P,\,(N)$ stands for the positive (negative) norm solutions. In the language of equation (\ref{equ11}) this technique means that one removes the effect of the background (flat space in that case) solutions. To illustrate this point let us
take the excited-de Sitter mode (\ref{equ8}) and calculate the spectrum with the auxiliary modes with de Sitter space as a background; then we have
$$ \langle{\phi^{2}}\rangle=\frac{1}{(2\pi)^{3}}\int
d^{3}{k}[\frac{1}{2ka^{2}}+\frac{H^{2}}{k^{3}}+\frac{a^{2}H^{4}}{8k^{5}}]-\frac{1}{(2\pi)^{3}}\int
d^{3}k[\frac{1}{2ka^{2}}+\frac{H^{2}}{2k^{3}}] $$ \b
=\frac{1}{2\pi^{2}}\int\frac{dk}{k}(\frac{H^{2}}{2}+\frac{a^{2}H^{4}}{8k^{2}}).
\e
The power spectrum is obtained by \cite{moh1} \b
P_{\phi}(k)=(\frac{H}{2\pi})^{2}\Big(1+\frac{1}{4}(\frac{H}{\Lambda})^{2}\Big),
\e which is scale-dependent and the order of correction is $(\frac{H}{\Lambda})^{2} $. Note that, similar correction has been obtained in \cite{Kaloper, Kemp}.

\subsection{Particle Production}
The concept of particle production in the early universe has been studied by several authors \cite{mot,bor}, however, the study of particle creation due to the gravitational field is commonly followed by two different methods. The first method defines the vacuum state at a particular instant of time as those which minimizes the energy by diagonalization of the Hamiltonian \cite{pav, gri}. On the other hand in the second method the vacuum is described as one for which the lowest-energy mode goes to zero smoothly in the far future and past represented by  $ "out" $ and $ "in" $ modes with a definite number of particles \cite{bri,ful}. However, this is not in the status of cosmology, because at least in the past, one cannot consider  the expansion of the universe as a smooth expansion. Then if there are no static $ "in" $ or $ "out" $ domains, this fact leads to the question of particle interpretation. In order to answer this question, a method of choosing those mode functions of the field equations that come in a sense closest to the Minkowski space-time limit are introduced. Physically, it corresponds to a concept of particle for which there is a minimum particle creation by the changing geometry \cite{bri}. However, here, we introduce a method of choosing the modes that come in a sense closest to the de Sitter space-time limit rather than flat space-time \cite{moh1,yus}. Therefore, we propose an excited-de Sitter mode (\ref{equ8}) as the fundamental mode during inflationary period that asymptotically approaches to de Sitter background mode. Accordingly the Bogoliubov coefficients can be computed and a straight calculation leads to the final phrase for the number of particles created in the $ k $ mode \cite{mij}
 \begin{equation} \label{equ13} \langle N \rangle=\frac{1}{4\omega_{k}(\eta)}|u'_{k}(\eta)|^{2}+\frac{\omega_{k}(\eta)}{4}|u_{k}(\eta)|^{2}-\frac{1}{2}. \end{equation}

Inspired by (\ref{equ11}) and (\ref{equ12}), the following definition for the spectrum of created particles, which caused by the perturbations of purely geometrical, may be defined
\begin{equation}
 \label{equ14}  \langle\Omega|:N(\eta):|\Omega\rangle=\langle\Omega|N(\eta)|\Omega\rangle-\langle0|N(\eta)|0\rangle,
\end{equation}
where $|0>$ is the vacuum of the background. At this stage two possibilities can be considered:
\begin{itemize}
\item \textbf{Flat space as a background}

To illustrate this viewpoint let us take the Bunch-Davies mode as a fundamental mode and calculate the spectrum of created particles when Minkowski space is supposed to be as a background; then we have:
$$
u_{k}^{BD}(\eta)=\frac{1}{\sqrt{k}}(1-\frac{i}{k\eta})e^{-ik\eta},\quad\quad  \omega_{k}^{2}(\eta)=k^{2}-\frac{2}{\eta^{2}} $$
$$ u_{k}^{Min}(\eta)=\frac{1}{\sqrt{k}}e^{-ik\eta},\quad\quad \omega_{k}^{2}(\eta)=k^{2}. $$
According to (\ref{equ13}) and (\ref{equ14}) and after doing some calculations one finds
$$ \langle\Omega|N(\eta)|\Omega\rangle_{BD}=-\frac{1}{2}+\frac{1}{4}|k^{2}\eta^{2}-2|^{1/2}(\frac{1}{k\eta}+\frac{1}{k^{3}\eta^{3}}) $$
\b +\frac{1}{4|k^{2}\eta^{2}-2|^{1/2}}(k\eta-\frac{1}{k\eta}+\frac{1}{k^{3}\eta^{3}}), \e
and
\b \langle0|N(\eta)|0\rangle_{Min}=0, \e
thus in this case the spectrum of created particles becomes
$$ \langle\Omega|:N(\eta):|\Omega\rangle=\langle\Omega|N(\eta)|\Omega\rangle_{BD}-\langle0|N(\eta)|0\rangle_{Min} $$
$$ =-\frac{1}{2}+\frac{1}{4}|k^{2}\eta^{2}-2|^{1/2}(\frac{1}{k\eta}+\frac{1}{k^{3}\eta^{3}}) $$
\b +\frac{1}{4|k^{2}\eta^{2}-2|^{1/2}}(k\eta-\frac{1}{k\eta}+\frac{1}{k^{3}\eta^{3}}), \e
which shows a growing behavior in the limit $ \eta\rightarrow 0 $. Also, the above spectrum show that in Minkowski space-time in the far past time limit, our result coincides to the standard results. Note that in \cite{per}, similar result has been obtained.

\item \textbf{de Sitter space as a background}

Now let us consider an approximate de Sitter mode as a fundamental mode during inflation. Therefore, we take the excited-de Sitter mode and calculate the spectrum of created particles in de Sitter background

$$
u_{k}^{exc}(\eta)=\frac{1}{\sqrt{k}}(1-\frac{i}{k\eta}-\frac{1}{2}(\frac{i}{k\eta})^{2})e^{-ik\eta},\quad\quad  \omega_{k}^{2}(\eta)=k^{2}-\frac{2}{\eta^{2}} $$
$$
u_{k}^{BD}(\eta)=\frac{1}{\sqrt{k}}(1-\frac{i}{k\eta})e^{-ik\eta},\quad\quad  \omega_{k}^{2}(\eta)=k^{2}-\frac{2}{\eta^{2}}. $$
After doing some straightforward algebra, one obtains
$$ \langle\Omega|N(\eta)|\Omega\rangle_{exc}=-\frac{1}{2}+\frac{1}{4}|k^{2}\eta^{2}-2|^{1/2}(\frac{1}{k\eta}+\frac{2}{k^{3}\eta^{3}}+\frac{1}{4k^{5}\eta^{5}}) $$
\b +\frac{1}{4|k^{2}\eta^{2}-2|^{1/2}}(k\eta+\frac{9}{4k^{3}\eta^{3}}+\frac{1}{k^{5}\eta^{5}}), \e
and
$$ \langle0|N(\eta)|0\rangle_{BD}=-\frac{1}{2}+\frac{1}{4}|k^{2}\eta^{2}-2|^{1/2}(\frac{1}{k\eta}+\frac{1}{k^{3}\eta^{3}}) $$
\b +\frac{1}{4|k^{2}\eta^{2}-2|^{1/2}}(k\eta-\frac{1}{k\eta}+\frac{1}{k^{3}\eta^{3}}). \e
Accordingly the spectrum of the created particles becomes as follows
$$ \langle\Omega|:N(\eta):|\Omega\rangle=\langle\Omega|N(\eta)|\Omega\rangle_{exc}-\langle0|N(\eta)|0\rangle_{BD} $$
$$ =\frac{1}{4}|k^{2}\eta^{2}-2|^{1/2}(\frac{1}{k^{3}\eta^{3}}+\frac{1}{4k^{5}\eta^{5}}) $$
\b +\frac{1}{4|k^{2}\eta^{2}-2|^{1/2}}(\frac{1}{k\eta}+\frac{5}{4k^{3}\eta^{3}}+\frac{1}{k^{5}\eta^{5}}), \e
which shows a growing behavior in the limit of $ \eta\rightarrow 0 $. Note that in \cite{per}, in the context of f(R)-gravity, the similar behavior has been discussed. Moreover, in the limit of $ \eta\rightarrow -\infty $ one obtains
\begin{equation}
 \label{equ15} \langle\Omega|N(\eta)|\Omega\rangle \approx \frac{1}{k^{2}\eta^{2}}, \end{equation}
which shows a reduction behavior in this limit which is in agreement with the standard results.\\
As the interesting result of this paper, we calculate the numerical order of the spectrum of created particles (\ref{equ15}) for the Danielsson cut-off time\cite{Dan1}, i.e. $ \eta_{0}=-\frac{\Lambda}{Hk}$, in the Planck time limit as
\b \langle\Omega|N(\eta)|\Omega\rangle \approx \frac{1}{k^{2}\eta_{0}^{2}}=(\frac{H}{\Lambda})^{2}=10^{-8}, \e
where $ H $ is the Hubble parameter and $ \Lambda $ is the fundamental energy scale of the theory during the inflation in the early universe.
\end{itemize}

\section{Conclusions}
In this work we had a look at the old problem of particle creation in de Sitter space-time. As a matter of fact the isometry of the de Sitter group rotates the solutions into a linear combination of the positive and negative ones. Therefore, the concept of particle and anti-particle becomes inexplicit. In this paper, we studied the cosmological particle production in the framework of general relativity by explicitly obtaining the general modes in an expanding background. By proposing an excited-de Sitter mode as the fundamental mode during the inflationary period, we calculated the spectrum of the created particles. \\
Inspired by the method of removing effects of the background space-time in a curved space-time in general and also the definition of two-point function in Krein space in particular, we offered a definition for spectrum of created particles which can be interpreted as the perturbation of the geometry. Then as an illustration, we calculated the spectrum of particles in de Sitter space as a background. We explicitly shown that during the inflation, massless particles can be produced by using the excited-de Sitter mode as the fundamental mode during inflationary period, so that in de Sitter and Minkowski space in the far past time limit our results reduces to the standard results. Finally, as the interesting result of this paper, the numerical magnitude of the spectrum of created particles, which caused by the perturbations of purely geometrical, in the Planck time limit is of order $10^{-8}$.\\

\noindent{\bf{Acknowlegements}}:
The authors would like to thank M.V. Takook and H. Pejhan for useful comments. This work has been supported by the Islamic Azad University, Qom Branch, Qom, Iran.

\end{document}